\documentstyle[11pt]{article}
\def\nz{\mbox{${\sf I \! N}$}}
\def\rz{\mbox{${\sf I \! R}$}}
\def\str{\mbox{$\scriptstyle\sf |$}}
 \def\cz{\mbox{$\sf C\!\!\!\mbox{\raisebox{0.45ex}{\str}}$}}
\newlength{\rrs}
 \newlength{\iis}
 \def\krz{\mbox{$\hspace{0.3\rrs}\mbox{{\scriptsize R}}\hspace{-1.15\rrs}%
  \mbox{{\scriptsize I}}\hspace{-1\iis}%
  \hspace{1.2\rrs}$}}
\def\d{{\rm d}}
\def\e{{\rm e}}
\def\i{\ifmmode{\rm i}\else\char"10\fi}

\def\ack
   {\vskip-\lastskip
    \vskip36pt plus12pt minus12pt
    \bigbreak
    \noindent{\large\bf Acknowledgments\par}
    \nobreak
    \bigskip
    \noindent}
\begin{document}
% Title and abstract
\title{Recent Developments in Supersymmetric Quantum Mechanics\thanks{Invited
talk presented at
``Frontiers in Theoretical Physics'', 15 - 21 December 1993,\newline
International Center for Physics and Applied Mathematics, Edirne, Turkey.
}}
\author{Georg Junker\\[10mm]
Institut f\"ur Theoretische Physik I,\\ Universit\"at
Erlangen-N\"urnberg, Staudtstr.\ 7,\\ D-91058 Erlangen, Germany\\[3mm]
\normalsize e-mail: junker@faupt101.physik.uni-erlangen.de}
\maketitle
\abstract{Some recent results in supersymmetric quantum mechanics are
presented. New semi-classical approximation formulas for Witten's realization
of supersymmetric quantum mechanics are discussed. Implications of the
supersymmetric structure of Pauli's Hamiltonian are also considered.
In particular, the paramagnetisation of a non-interacting electron gas is
related to a modified version of Witten's index.
We also show that the supersymmetry in this system
provides a new counter example for the paramagnetic conjecture.}
% Section 1
\setcounter{equation}{0}
\section{Introduction}
For more than two decades the idea of supersymmetry (SUSY) has been inspiring
physicists. Originally, SUSY has been introduced to formulate a quantum field
theory which unifies bosons and fermions \cite{SUSY1,SUSY2}. It is a fact
that bosons and fermions are clearly distinguishable and SUSY is apparently
broken in our present environment. Nevertheless, the
SUSY idea has led to new insights in the studies of nuclear physics, condensed
matter physics, statistical physics and mathematical physics \cite{KC85}. In
particular, supersymmetric quantum mechanics, which originally has been
introduced in 1976 by Nicolai \cite{N76} and rediscovered in 1981 by Witten
\cite{W81}, nowadays attracts much attention. In this report we will present
some of the recent developments in SUSY quantum mechanics.

In Section 2 we will give the definition of SUSY in a quantum mechanical
system. Section 3 deals with new developments in Witten's ($N=2$)-realization
of the SUSY algebra, which essentially is a pair of one-dimensional quantum
systems. In particular, the SUSY-inspired semi-classical quantization
condition in this realization, which differs from the usual WKB condition,
has led to remarkable results. These are presented in the Sections 3.1 and
3.2. In Section 4 we will consider another realization of the SUSY algebra.
This ($N=1$)-realization leads to the Pauli Hamiltonian of an electron in an
arbitray magnetic field. We will discuss, in Section 4.1, the implications of
this SUSY structure on the paramagnetism of a two-dimensional non-interacting
electron gas. Section 4.3 deals with the three-dimensional case. Here we
present a new counterexample to the paramagnetic conjecture.
% Section 2
\setcounter{equation}{0}
\section{Supersymmetric quantum mechanics}
Following  Nicolai \cite{N76} we call a quantum mechanical system
characterized by a Hamiltonian $H$ acting in some Hilbert space ${\cal H}$
{\em supersymmetric} if there exist selfadjoint operators
$Q_{i}=Q_{i}^{\dagger}$, $i=1,2,\ldots,N$, called {\em supercharges},
which also act on states in ${\cal H}$ and fulfill the following SUSY algebra:
\begin{equation}
\begin{array}{c}
\{Q_{i},Q_{j}\}:=Q_{i}Q_{j}+Q_{j}Q_{i}=H\delta _{ij},\\[2mm]
[Q_{i},H]:=Q_{i}H-HQ_{i}=0,~~~i,j=1,2,\ldots,N.
\end{array}
\end{equation}

In addition to this definition,
it may be convenient to postulate the existence of a
selfadjoint operator $P=P^{\dagger}$, called {\em Witten operator} or {\em
Witten parity}, which anticommutes with the supercharges, and therefore
commutes with the Hamiltonian, and whose square is equal to the identity
[6-8]:
\begin{equation}
\{Q_{i},P\}=0,~~~[H,P]=0,~~~P^{2}=1.
\end{equation}
This operator allows to introduce the notion of ``bosonic'' and ``fermionic''
states independently of an underlying space-time symmetry.
Actually, the Witten parity is sometimes also written in the form $P=(-1)^{F}$
where $F$ is the fermion-number operator. Therefore, eigenstates of $P$ with
eigenvalue $-1$ are called ``fermions'' and those with eigenvalue $+1$ are
said to be ``bosons''. It should be stressed, that these kinds of bosons and
fermions may in general not be interpreted as particles with integer and
half-integer spin, respectively. To make this point more clear let us
introduce a bosonic subspace ${\cal H}_{\rm B}$ and a fermionic subspace
${\cal H}_{\rm F}$:
\begin{equation}
{\cal H}_{\rm B}:=\{|\Psi \rangle\in{\cal H}\,|\,P|\Psi \rangle=+|\Psi
\rangle\},~~~
{\cal H}_{\rm F}:=\{|\Psi \rangle\in{\cal H}\,|\,P|\Psi \rangle=-|\Psi
\rangle\}.
\end{equation}
Hence, any state $|\Psi \rangle\in{\cal H}$ can be decomposed into its bosonic
and fermionic components as follows:
\begin{equation}
|\Psi \rangle=
\left(\begin{array}{c}|\psi_{\rm B}\rangle\\ |\psi_{\rm
F}\rangle\end{array}\right).
\end{equation}
The Hilbert space may be
written as a product space ${\cal H}={\cal H}_{0}\otimes \cz^{2}$. It is
obvious, that the Witten operator is represented by the third Pauli matrix
$\sigma _{3}$:
\begin{equation}
P={\bf 1}\otimes\sigma _{3}=\left(\begin{array}{rr}{\bf 1}&0\\0&-{\bf
1}\end{array}\right).
\end{equation}
Therefore, it would be more appropriate to use the notion ``spin-up'' and
``spin-down'' states (of a fictitious spin-$\frac{1}{2}$-particle)
instead of bosonic and fermionic states, respectively.
Having in addition only cartesian degrees of freedom
${\cal H}_{0}$ is given by the space of square-integrable functions over the
$d$-dimensional Euclidean space $\rz^d$, ${\cal H}_{0}={\cal L}^{2}(\rz^{d})$,
$d\in\nz$.
As $P$ anticommutes with all supercharges the latter are necessarily of the
form
\begin{equation}
\begin{array}{rcl}
Q_{1}&:=&\displaystyle\frac{\i}{\sqrt{2}}
\left(A_{1}^{\dagger}\otimes\sigma _{+}-A_{1}\otimes\sigma _{-}\right)=
\frac{\i}{\sqrt{2}}
\left(\begin{array}{rr}0_{~}&A_{1}^{\dagger}\\-
A_{1}&0_{~}\end{array}\right)\\[6mm]
Q_{2}&:=&\displaystyle\frac{1}{\sqrt{2}}
\left(A_{2}^{\dagger}\otimes\sigma _{+}+A_{2}\otimes\sigma _{-}\right)=
\frac{1}{\sqrt{2}}
\left(\begin{array}{rr}0_{~}&A_{2}^{\dagger}\\A_{2}&0_{~}\end{array}\right)\\
&\vdots&
\end{array}
\end{equation}
where $A_{i}$, $i=1,2,\ldots,N$, are operators acting on states in
${\cal H}_{0}$ and
$\sigma_{\pm}:=(\sigma_{1}\pm\i\sigma_{2})/2$ are the usual raising and
lowering operators for the eigenvalues of $\sigma _{3}$. It is
clear that the supercharges transform spin-up states (bosons) into spin-down
states (fermions) and vice versa. This is the analogue of the well-known SUSY
transformations in four-dimensional SUSY field theory.

The supersymmetry (2.1) has also implications on the spectral properties of
the Hamiltonian $H$. First of all, we note that $H=2Q_{i}^{2}\geq 0$. That is,
the Hamiltonian has only non-negative eigenvalues. Let us suppose that
$|\Psi _{r}\rangle$ is an eigenstate of $H$ with positive eigenvalue $E_{r}>0$.
Then it follows immediately from the algebra (2.1) that
\begin{equation}
|\tilde{\Psi }_{r}\rangle=\left(2/E_{r}\right)^{1/2}\,Q_{i}|\Psi _{r}\rangle,
{}~~i=1,2,\ldots,N,
\end{equation}
is also an
eigenstate with the same positive eigenvalue. Hence, all positive-energy
eigenstates occur in spin-up (boson) spin-down (fermion) pairs. This, however,
is in general not true for possible zero-energy eigenstates. If the
groundstate energy of $H$ is equal to zero, that is, there exists a state
$|\Psi _{0}\rangle\in{\cal H}_{0}\otimes\cz^2$ such that
\begin{equation}
H|\Psi _{0}\rangle=0,
\end{equation}
then SUSY is said to be a good symmetry. This is, because (2.8) implies that
the groundstate is invariant under SUSY transformations,
$Q_{i}|\Psi _{0}\rangle=0$. If the groundstate energy of $H$ is strictly
positive then SUSY is said to be broken. Figure 1 shows typical spectra
for good and broken SUSY, respectively.

In order to decide whether SUSY is a good symmetry Witten \cite{W82}
introduced the following trace
\begin{equation}
\Delta (\beta ):={\rm tr}\,\left(P\e^{-\beta H}\right),~~~\beta >0.
\end{equation}
This quantity is called {\em Witten index}.
For a pure point spectrum of $H$ this
index is the difference of the number of spin-up states $(\uparrow)$
and spin-down states $(\downarrow)$ with zero energy:
\begin{equation}
\Delta (\beta )=N_{\uparrow}(E=0)-N_{\downarrow}(E=0).
\end{equation}
Hence, $\Delta (\beta )\neq 0$ implies good SUSY. Obviously, the Witten index
is independent of $\beta $. Indeed, the factor $\e^{-\beta H}$ in (2.9) has
only been introduced \cite{Comment1} for regularisation of the trace.
The contributions of the positive-energy
eigenstates cancel due to the pairwise degeneracy mentioned above.
For a continuous spectrum this is not the case as the spectral densities for
the spin-up and spin-down states are in general different. Here, the Witten
index becomes $\beta $ dependent \cite{FC85}.
% Section 3
\setcounter{equation}{0}
\section{Witten's realization}
In this Section we will consider a particular $(N=2)$-realization of the SUSY
algebra (2.1) in the Hilbert space ${\cal H}={\cal L}^{2}(\rz)\otimes\cz^2$.
The two supercharges are given by (2.6) with
\begin{equation}
A_{1}=A_{2}\equiv A:=\frac{\i}{\sqrt{2m}}\,p +\phi(x).
\end{equation}
Here $p$ and $x$ denote the usual momentum and position operators in ${\cal
H}_{0}={\cal L}^{2}(\rz)$ and $\phi:\rz\mapsto\rz$ is a piecewise continuously
differentiable function called {\em SUSY potential}. The above realization
has first been used by Witten \cite{W82} for studying the dynamical
SUSY-breaking mechanism.

In terms of the SUSY potential $\phi$ the Hamiltonian reads
$(\phi':=\d\phi/\d x)$
\begin{equation}
H=\left(\frac{p^2}{2m}+\phi^2(x)\right)\otimes{\bf 1}-
\frac{\hbar}{\sqrt{2m}}\,\phi'(x)\otimes\sigma _{3}
\end{equation}
which describes a spin-$\frac{1}{2}$ point particle of mass $m>0$ moving along
the Euclidean line under the influence of the external scalar potential
$\phi^2$ and a magnetic field which is orthogonal to this axis and whose
strength depends linearly on the slope of the SUSY potential. Obviously, the
Hamiltonian is diagonal,
\begin{equation}
H=\left(\begin{array}{cc}H_{-}&0\\0&H_{+}\end{array}\right),
\end{equation}
and, therefore, the spin-up (bosonic) Hamiltonian $H_{-}$ and spin-down
(fermionic) Hamiltonian $H_{+}$, which are defined as follows,
\begin{equation}
\begin{array}{rcl}
H_{-}&:=&A^{\dagger}A=\displaystyle
\frac{p^2}{2m}+\phi^2(x)-\frac{\hbar}{\sqrt{2m}}\,\phi'(x),\\[2mm]
H_{+}&:=&AA^{\dagger}=\displaystyle
\frac{p^2}{2m}+\phi^2(x)+\frac{\hbar}{\sqrt{2m}}\,\phi'(x) ,
\end{array}
\end{equation}
can be considered separately. This allows the alternative interpretation that
Witten's realization (3.1) characterizes two non-interacting point particles
of equal
mass $m$ moving along the real line under the influence of the external scalar
potentials $V_{-}$ and $V_{+}$, respectively, where
\begin{equation}
V_{\pm}(x):=\phi^2(x)\pm\frac{\hbar}{\sqrt{2m}}\,\phi'(x).
\end{equation}
Let us denote by $|\psi ^{+}_{r}\rangle$ and $|\psi ^{-}_{r}\rangle$ the
eigenstates of $H_{+}$ and $H_{-}$, respectively, for the same positive
eigenvalue $E_{r}>0$:
\begin{equation}
H_{\pm}|\psi _{r}^{\pm}\rangle=E_{r}|\psi _{r}^{\pm}\rangle.
\end{equation}
Then the SUSY transformation (2.7) implies the relations
\begin{equation}
|\psi _{r}^{+}\rangle=\frac{1}{\sqrt{E_{r}}}\,A|\psi _{r}^{-}\rangle,~~~
|\psi _{r}^{-}\rangle=\frac{1}{\sqrt{E_{r}}}\,A^{\dagger}|\psi_{r}^{+}\rangle.
\end{equation}

It should be noted that under the replacement $\phi\to -\phi$ the roles of the
two Hamiltonians $H_{+}$ and $H_{-}$ are interchanged, that is $H_{\pm}\to
H_{\mp}$. Hence, the sign of the SUSY potential may be fixed by some
convention. The usual convention is that the groundstate energy belongs to
a spin-up state (boson):
\begin{equation}
{\rm inf\,spec\,}(H)={\rm inf\,spec\,}(H_{-}).
\end{equation}

For good SUSY there exists in addition to the states (3.6) the ground
state $|\psi _{0}\rangle$ defined in accordance with our convention (3.8) by
\begin{equation}
H_{-}|\psi _{0}\rangle=0~~~\Longleftrightarrow~~~A|\psi _{0}\rangle=0.
\end{equation}
In the coordinate representation this state reads
\begin{equation}
\langle x|\psi _{0}\rangle\equiv \psi _{0}(x)=
\psi _{0}(0)\exp\left\{-\frac{\sqrt{2m}}{\hbar}\int\limits_{0}^{x}\d
z\,\phi(z)\right\}
\end{equation}
which has to be square-integrable for SUSY to be a good symmetry.
This requirement puts conditions on the SUSY potential $\phi$. In
Figure 2 we show typical $\phi$'s which lead to good and broken SUSY,
respectively.
For continuous SUSY potentials the function $\phi$ must have an
odd number of zeros (counted with their multiplicity) for SUSY to be good. A
continuous SUSY potential with an even number of zeros necessarily leads to a
broken SUSY as (3.10) will be not square-integrable.
Consequently, if $\phi$ has a well-defined parity,
an odd $\phi$ leads to good SUSY, whereas an even $\phi$ breaks SUSY
\cite{IJ92}:
\begin{equation}
\begin{array}{lll}
\phi(-x)=-\phi(x)&\Rightarrow&V_{\pm}(-x)=V_{\pm}(x)~~~({\rm
SUSY~and~parity~are~good}),\\[2mm]
\phi(-x)=\phi(x)&\Rightarrow&V_{\pm}(-x)\neq V_{\pm}(x)~~~({\rm
SUSY~and~parity~are~broken}).
\end{array}
\end{equation}
The spectra of the two Hamiltonians (3.4) are related as follows:
\begin{equation}
\begin{array}{rll}
{\rm spec}(H_{-})\backslash\{0\}&={\rm spec}(H_{+})&~~~({\rm
good~SUSY}),\\[2mm]
{\rm spec}(H_{-})&={\rm spec}(H_{+})&~~~({\rm broken~SUSY}).
\end{array}
\end{equation}

Finally, let us mention that any one-dimensional standard
problem, that is, the Schr\"odinger problem (in ${\cal H}_{0}$) for a point
mass $m$ in a scalar potential $U(x)$, can be put into this SUSY formalism.
For example, knowing one (not necessarily square-integrable) solution
$\varphi (x)$ of the Schr\"odinger-like equation
\begin{equation}
-\frac{\hbar^{2}}{2m}\,\varphi ''(x)+U(x)\varphi (x)=\varepsilon \varphi (x)
\end{equation}
a family of solutions of the generalized Riccati equation \cite{S85}
\begin{equation}
\phi^{2}(x)-\frac{\hbar}{2m}\,\phi'(x)=U(x)-\varepsilon
\end{equation}
is given by
\begin{equation}
\phi(x):=-\frac{\hbar}{\sqrt{2m}}\left(\frac{\varphi '(x)}{\varphi(x)}+
\frac{\varphi ^{-2}(x)}{\lambda +\int\d x\,\varphi ^{-2}(x)}\right).
\end{equation}
It should be noted that not all parameters $\lambda \in \rz$ will lead to an
admissible SUSY potential $\phi$ such that $H_{\pm}$ will be selfadjoint on
${\cal H}_{0}$. However, for $\lambda \to\infty $,
$\varepsilon $ being the groundstate energy of the Schr\"odinger problem
(3.13) and $\varphi (x)$ being the corresponding normalized groundstate
wavefunction, the SUSY potential (3.15) takes the form $\phi(x)=-
(\hbar/\sqrt{2m})\varphi '(x)/\varphi (x)$ and will be admissible.
\subsection{Semi-classical approximations}
Recently, Witten's realization (3.1) of SUSY quantum mechanics has attracted
much
attention. This is mainly due to a new semi-classical approximation which leads
for such SUSY systems to better results then the old WKB approximation does.
In fact, for basically all exactly solvable problems the new approximation
leads to exact energy spectra. This is in contrast to the WKB approach which
in general requires ad hoc Langer-like modifications for those systems in order
to yield the exact spectrum. The new approximation formula, which can be
derived from Feynman's path integral using a modified
stationary-action-principle \cite{IJ92,IJ93,IJ94}, reads
\begin{equation}
\int\limits_{q_{\rm L}}^{q_{\rm R}}\d x\sqrt{2m[E-\phi^{2}(x)]}=
\left(n+\frac{1}{2}-\frac{a(q_{\rm L})-a(q_{\rm R})}{2\pi }\right)\pi \hbar,
{}~~n=0,1,2\ldots,
\end{equation}
where
\begin{equation}
a(x):={\rm arcsin}\Bigl(\phi(x)/\sqrt{E}\Bigr).
\end{equation}
The left and right turning points $q_{\rm L}\leq q_{\rm R}$ are defined by
$\phi^{2}(q_{\rm L})=E=\phi^{2}(q_{\rm R})$. We note that this quantization
condition holds for both potentials $V_{\pm}$. This is not the case for the WKB
formula which reads
\begin{equation}
\int\limits_{x_{\rm L}}^{x_{\rm R}}\d x\sqrt{2m[E-V_{\pm}(x)]}=\textstyle
\left(n+\frac{1}{2}\right)\pi \hbar,
{}~~n=0,1,2\ldots~.
\end{equation}
Here the turning points are given by $V_{\pm}(x_{\rm L})=E=V_{\pm}(x_{\rm R})$.
Formally, the previous expression (3.16) can be derived from the WKB formula by
inserting the explicit form (3.5) of $V_{\pm}$ and treating
$\phi$ to be independent of $\hbar$ (eq.\ (3.15) clearly shows that this is in
general not the case). An expansion in $\hbar$, see for example
\cite{E86,IJS93}, then leads to the desired result (3.16).

There are two possible solutions for the roots of the turning-point-condition
$\phi^{2}(q_{\rm L})=E=\phi^{2}(q_{\rm R})$ which can be related to good and
broken SUSY. The first possibility $\phi(q_{\rm L})=-\phi(q_{\rm R})=
\pm\sqrt{E}$ corresponds to good SUSY with the upper sign for $H_{+}$ and the
lower one for $H_{-}$ in accordance with the convention (3.8)
(see also Figure 2). The second possibility, $\phi(q_{\rm L})=\phi(q_{\rm R})=
\pm\sqrt{E}$, is related to a broken SUSY. More explicitly, formula (3.16)
can be put into the form
\begin{equation}
\int\limits_{q_{\rm L}}^{q_{\rm R}}\d x\sqrt{2m[E-\phi^{2}(x)]}=
(n+\nu )\pi \hbar,~~~n=0,1,2\ldots,
\end{equation}
where
\begin{equation}
\nu :=\left\{\begin{array}{ll}
0~~~{\rm for}~H_{-}~{\rm and~good~SUSY}\\
1~~~{\rm for}~H_{+}~{\rm and~good~SUSY}\\
\frac{1}{2}~~~{\rm for}~H_{\pm}~{\rm and~broken~SUSY}
\end{array}\right. .
\end{equation}
For good SUSY ($\nu=0,1$) this semi-classical quantization condition has first
been suggested by Comtet, Bandrauk and Campbell \cite{CBC85} using the formal
$\hbar$ expansion mentioned above.

Let us note that this new quantization condition is in accordance with
the spectral relations (3.12).
This is generally not true for the WKB approximation.
Furthermore, for good SUSY the condition (3.19) leads to the exact groundstate
energy $E=0$ of $H_{-}$. Thus being exact for the ground state and
(as any semi-classical approximation) a good approximation for
large quantum numbers $n$ it may be conjectured that the SUSY approximation
formula (3.19) might lead to estimates for energy eigenvalues with
intermediate quantum numbers which are better than those of the WKB
approximation (3.18). Indeed, for all so-called shape-invariant potentials
\cite{G83} (the shape-invariance condition is identical to the
factorizability of Schr\"odinger, Infeld and Hull \cite{IH51})
it has been found that the SUSY formula (3.19) does provide
the exact spectrum. The shape-invariant potentials can be put into
three classes \cite{G83}:\\[2mm]
Class 1:
\begin{equation}
\begin{array}{l}
\displaystyle
\phi_{1}(x):=\frac{\hbar}{\sqrt{2m}}\Bigl( af_{1}(x)+b\Bigr)~~~{\rm with}~~~
f'_{1}(x)=pf_{1}^{2}(x)+qf_{1}(x)+r,\\[2mm]
\displaystyle
V^{(1)}_{\pm}(x)=\frac{\hbar^2}{2m}\left[
a(a\pm p)f_{1}^{2}(x)+a(2b\pm q) f_{1}(x)+(b^{2}\pm ar)\right].
\end{array}
\end{equation}
Class 2:
\begin{equation}
\begin{array}{l}
\displaystyle
\phi_{2}(x):=\frac{\hbar}{\sqrt{2m}}\Bigl(af_{2}(x)+b/f_{2}(x)\Bigr)
{}~~{\rm with}~~~f'_{2}(x)=pf_{2}^{2}(x)+q,\\[2mm]
\displaystyle
V^{(2)}_{\pm}(x)=\frac{\hbar^2}{2m}\left[
a(a\pm p)f_{2}^{2}(x)+b(b\mp q)/f^{2}_{1}(x)+2ab\pm(aq-bp)\right].
\end{array}
\end{equation}
Class 3:
\begin{equation}
\begin{array}{l}
\displaystyle
\phi_{3}(x):=\frac{\hbar}{\sqrt{2m}}\left(a+b\sqrt{pf_{3}^{2}(x)+q}\,\right)/f_{3}(x)
{}~~{\rm with}~~~f'_{3}(x)=\sqrt{pf_{3}^{2}(x)+q},\\[2mm]
\displaystyle
V^{(3)}_{\pm}(x)=\frac{\hbar^2}{2m}\left[\frac{a^{2}+pq(b\mp1)}{f_{3}^{2}(x)}+
\frac{\sqrt{pf_{3}^{2}(x)+q}}{f_{3}^{2}(x)}\,a(2b\mp1)+b^{2}p\right].
\end{array}
\end{equation}
Here $a,b,p,q,r\in\rz$ are arbitrary potential parameters.
Shape invariance means that the potential $V^{(i)}_{+}$ can be written in
terms of $V^{(i)}_{-}$ after appropriate redefinitions of the potential
parameters. For example, in
classes 1 and 2 we may perform the reparameterization $(p,q,r)\to(-p,-q,-r)$
which results in $V^{(1,2)}_{-}\leftrightarrow V^{(1,2)}_{+}$.
We note that the potentials belonging to class 2 posses additional
reparameterisation invariances:
\begin{eqnarray}
a\to -(a\pm p)&~~~\Rightarrow~~~&\phi_{2}(x)\to -(a\pm p)f_{2}(x)+b/f_{2}(x)
\\[2mm]
b\to -(b\mp q)&~~~\Rightarrow~~~&\phi_{2}(x) \to af_{2}(x)-(b\mp q)/f_{2}(x) .
\end{eqnarray}
In both cases the full potentials are only shifted by an additional constant,
$V^{(2)}_{\pm}(x)\to V^{(2)}_{\pm}(x)+{\rm const}$.
The particular form of the SUSY potential $\phi_{2}$ allows one to choose
parameters such that SUSY will either be good or broken. These
facts have first been pointed out by Suparmi \cite{S92} (see also
\cite{IJS93,DGKPS93}). Let us also mention that the shape-invariance
condition,
which is equivalent to the factorization condition, implies also dynamical
symmetries in addition to SUSY. For a recent discussion of the embedding of
this SUSY structure into the dynamical group structure of shape-invariant
potentials we refer to the work by Barut and Roy \cite{BR92}.

As already pointed out, the interesting observation
is that the new approximation formula (3.19) leads to the exact spectrum of
all those shape-invariant potentials.
Whereas the classes 1 and 3 allow only for a good SUSY, class 2 also
provides broken SUSY examples \cite{IJ93,IJS93}.
Let us consider the example of the radial harmonic oscillator which belongs
to class 2 with $f_{2}(x):=x$, $a:=m\omega /\hbar>0$, $q:=1$ and $p:=0$. Here
${\cal H}_{0}={\cal L}^2(\rz^+)$ with Dirichlet boundary condition at $x=0$.
The choice $b:=-l-1\leq -1$ leads to the SUSY potential
\begin{equation}
\phi(x)=\sqrt{\frac{m}{2}}\omega x-\frac{(l+1)\hbar}{\sqrt{2m}\,x}
\end{equation}
with good SUSY. The corresponding full potentials read
\begin{equation}
\begin{array}{rcl}
V^{(2)}_{-}(x)&=&\displaystyle\frac{m}{2}\omega
^2x^2+\frac{l(l+1)\hbar^2}{2mx^2}
\textstyle -\hbar\omega (l+\frac{3}{2}),\\[4mm]
V^{(2)}_{+}(x)&=&\displaystyle\frac{m}{2}\omega
^2x^2+\frac{(l+1)(l+2)\hbar^2}{2mx^2}
\textstyle -\hbar\omega (l+\frac{1}{2}) .
\end{array}
\end{equation}
Application of the semi-classical quantization condition (3.19) gives
\cite{S92}
\begin{equation}
E_{n}=2\hbar\omega (n+\nu ).
\end{equation}
This is indeed the exact spectrum for (3.27). For the alternative choice
$b:=l\geq 0$ we have
\begin{equation}
\phi(x)=\sqrt{\frac{m}{2}}\omega x+\frac{l\hbar}{\sqrt{2m}\,x}
\end{equation}
which gives a broken SUSY. The full potentials read in this case
\begin{equation}
\begin{array}{rcl}
V^{(2)}_{-}(x)&=&\displaystyle\frac{m}{2}\omega
^2x^2+\frac{l(l+1)\hbar^2}{2mx^2}
\textstyle +\hbar\omega (l-\frac{1}{2}),\\[4mm]
V^{(2)}_{+}(x)&=&\displaystyle\frac{m}{2}\omega
^2x^2+\frac{l(l-1)\hbar^2}{2mx^2}
\textstyle +\hbar\omega (l+\frac{1}{2}).
\end{array}
\end{equation}
Here, the formula (3.19) has to be applied with $\nu =1/2$. Again the exact
spectrum
\begin{equation}
E_{n}=\hbar\omega (2n+2l+1)
\end{equation}
can be obtained \cite{S92}. We finally note that the WKB approximation (3.18)
provides these exact results only after making the Langer modification
$l(l+1)\to (l+1/2)^2$.
\subsection{Numerical results for a class of power potentials}
Being superior to the WKB approximation in the case of shape-invariant
problems, there naturally arises the question whether the new approximation
(3.19) also provides better estimates for not exactly solvable systems. To
answer this questions we have investigated a class of power SUSY potentials
of the form \cite{IJ94}
\begin{equation}
\phi(x):=\frac{\hbar a}{\sqrt{2m}} x^{d}~~~{\rm for}~~~x \geq 0.
\end{equation}
Here $a>0$ and $d=1,2,\ldots$, are free parameters. Note that the above
definition is only for the positive Euclidean half-line.
For $x<0$ we may define
the SUSY potential either through an antisymmetric or symmetric continuation
which leads to a good and broken SUSY, respectively:
\begin{equation}
\begin{array}{ll}
\phi(-x):=-\phi(x)&~~~{\rm for~good~SUSY},\\
\phi(-x):=\phi(x)&~~~{\rm for~broken~SUSY}.
\end{array}
\end{equation}
The full potentials read
\begin{equation}
\begin{array}{lll}
V_{\pm}(x)&=\displaystyle
\frac{\hbar^2}{2m}\left(a^2x^{2d}\pm ad\,|x|^{d-1}\right)=V_{\pm}(-x)
&~~~{\rm for~good~SUSY},\\[2mm]
V_{\pm}(x)&=\displaystyle
\frac{\hbar^2}{2m}\left(a^2x^{2d}\pm ad\,x|x|^{d-2}\right)=V_{\mp}(-x)
&~~~{\rm for~broken~SUSY}.
\end{array}
\end{equation}

For the SUSY potential (3.32) the new semi-classical formula (3.19) can be
evaluated analytically with the result
\begin{equation}
E_{n}=\frac{\hbar^2a^2}{2m}\left[
\frac{\Gamma\left(\frac{3d+1}{2d}\right)}{\Gamma\left(\frac{2d+1}{2d}\right)}
\frac{\sqrt{\pi }}{a}(n+\nu )\right]^{2d/(d+1)}.
\end{equation}
These energy eigenvalues are in general not exact. For arbitrary $d$ only
the case $n=\nu=0$ in (3.35) gives the exact result, which is the
groundstate energy for good SUSY. The associated wavefunction (3.10)
reads explicitly
\begin{equation}
\psi _{0}(x)=\psi _{0}(0)\exp\left\{-\frac{a}{d+1}|x|^{d+1}\right\}.
\end{equation}
Another case where (3.35) becomes exact is the limit $d\to\infty $:
\begin{equation}
\lim_{d\to\infty }E_{n}=\frac{\hbar^2\pi ^2}{8m}\,(n+\nu )^2.
\end{equation}
That this is indeed the exact spectrum can be seen by realizing that (3.34)
become infinite square-well potentials:
\begin{equation}
\lim_{d\to\infty }V_{\pm}(x)=\left\{\begin{array}{lll}
0~~~&{\rm for}~~~&|x|<1\\ \infty ~~~&{\rm for}~~~&|x|>1\end{array}\right.
\end{equation}
This limit has to be taken with some care. Actually, the Hilbert space changes
in this limit, too: ${\cal L}^2(\rz)\to{\cal L}^2([-1,1])$. Hence, now one has
to specify boundary conditions at $x=\pm 1$ in order to have a well-defined
problem. The type of boundary conditions which has to be chosen is related to
the requirement that the SUSY structure, that is, good or broken SUSY,
is conserved after performing the limit $d\to\infty $.
It is obvious that for this one has to impose, for good
SUSY, Neumann conditions at $x=\pm 1$ for $V_{-}$ and Dirichlet condition at
$x=\pm 1$ for $V_{+}$. Thus parity as well as SUSY remain to be good
symmetries. For broken SUSY and broken parity we have to choose for $V_{-}$
Dirichlet conditions at $x=-1$ and Neumann conditions at $x=1$, and for
$V_{+}$ vice versa. Let us note that the spectrum (3.35) also appears in
the contribution of Uzes, Barut and Kapuscik \cite{UBK94} to these
proceedings, where they discuss the quantization of a constrained
complex non-linear oscillator. Indeed, the Hamiltonian (3.2) in Witten's
realization can be derived from a more general SUSY Hamiltonian with
second-class constraints \cite{N91}.

For finite $d$ we have compared the eigenvalues (3.35) and those obtained via
the WKB formula (3.18) with the numerically exact eigenvalues of the
Schr\"odinger equation for the full potentials (3.34).
In Figure 3 and 4 we plot the relative deviations (in $\%$)
\begin{equation}
\Delta:=(E_{\rm exact}-E_{\rm approx})/E_{\rm exact}
\end{equation}
for the cases $d=2,3$ and for good and broken SUSY, respectively.
For the numerical evaluation we used units such that $m=\hbar=a=1$. As
can clearly be seen in these Figures, the new approximation formula
provides estimates for the eigenvalues which are above of those of the WKB
approximation. Except for the first few eigenvalues for good SUSY we
made the observation that the new formula (3.19) always overestimates the
exact energy value whereas the WKB formula (3.18) gives underestimations. We
have made this observation not only for the class of power potentials
characterized by (3.32) but for all systems investigated so far.
In Figure 5 we present results for $\phi(x)=\sinh(x)$ and $\phi(x)=\cosh(x)$
with good and broken SUSY, respectively, showing the same tendency
\cite{IJ94}.

We conclude that the new approximation (3.19) in general does not provide
estimates better than those obtained from the WKB approximation (3.18).
However, combining both by taking the mean value will give much better
estimates \cite{IJ94}.
% Section 4
\setcounter{equation}{0}
\section{Supersymmetry in the Pauli Hamiltonian}
In this Section we will consider a ($N=1$)-realization of the algebra (2.1) in
the Hilbert space ${\cal H}={\cal L}^2(\rz^2)\otimes\cz^2$. Here the
supercharge $Q\equiv Q_{1}$ is defined by
\begin{equation}
Q:=\frac{1}{\sqrt{4m}}\sum_{i=1}^2\left[\left(p_{i}-
\frac{e}{c}\,a_{i}(x,y)\right)
\otimes\sigma _{i}\right]
\end{equation}
where $e$ denotes the charge of a point mass $m$, $c$ is the speed of light,
and
$a_{i}:\rz^2\mapsto\rz$ are components of an external vector potential
characterizing a magnetic field of strength $B(x,y):=\partial_{x}a_{2}(x,y)-
\partial_{y}a_{1}(x,y)$ which is perpendicular to the $x$-$y$-plane. The
corresponding SUSY Hamiltonian $H=2Q^2$ explicitly reads
\begin{equation}
H=\frac{1}{2m}\sum _{i=1}^2
\left(p_{i}-\frac{e}{c}\,a_{i}(x,y)\right)^2\otimes{\bf 1}-
\mu_{\rm B} B(x,y)\otimes\sigma _{3}
\end{equation}
where $\mu_{\rm B} :=e\hbar/(2mc)$ is the Bohr magneton. Hence, the
Hamiltonian (4.2) is identical with the Pauli Hamiltonian for a two-dimensional
electron of charge $e$, mass $m$ and gyromagnetic ratio equal to two. This
realization of the SUSY algebra (2.1) has first been given by deCrombrugge and
Rittenberg \cite{CR83}. However, the observation that the Pauli Hamiltonian
factorizes has already been made by Kramers \cite{K38}. In this realization
the bosonic and fermionic states are the usual spin-up and spin-down states of
the electron. Let us also note that there is a close relation between the
supercharge (4.1) and the Dirac operator. For more details we refer to the
excellent book by Thaller \cite{T92} which makes extensive use of SUSY in the
discussion of the Dirac equation.

Due to a famous result of Aharonov and Casher \cite{AC78}
(see also \cite{CFKS87,T92}) the groundstate energy of the Hamiltonian (4.2) is
zero and belongs either to spin-up or spin-down states. Therefore,
SUSY is a good symmetry for an arbitrary magnetic field strength $B$.
The nature of the groundstates can, as before, be fixed by some convention
which essentially corresponds to an
appropriate choice of the overall sign of $B$. By choosing this sign
such that the net magnetic flux through the $x$-$y$-plane is positive the
groundstate energy belongs to spin-up states only. The degeneracy of the
groundstate is given by
\begin{equation}
g:=\left[\!\!\left[\frac{1}{\Phi _{0}}\int\limits_{\krz^2}\d x\d y\,B(x,y)
\right]\!\!\right]
\end{equation}
where $\Phi _{0}:=2\pi \hbar c/e$ is the flux quantum and $[\![z]\!]$ denotes
the largest integer which is strictly less than $z$.
For a pure point spectrum (this occurs for example in the case of a constant
magnetic field \cite{Comment2})
the degeneracy $g$ is identical to the Witten index (2.9):
\begin{equation}
g=\Delta (\beta )=N_{\uparrow}(E=0).
\end{equation}
The equality of the two right-hand-sides of (4.3) and (4.4) is a physical
realization of the Atiyah-Singer index theorem relating the algebraic index of
an operator with its topological index.
\subsection{Paramagnetism of the two-dimensional electron gas}
The SUSY of the Pauli Hamiltonian (4.2) allows to study its paramagnetic
properties for arbitrary magnetic fields $B$. Let us first assume that
the magnetic field is such that (4.2) has a pure point spectrum. Then, because
of SUSY, we know that all subspaces with a strictly positive energy
eigenvalue contain pairs of spin-up--spin-down states.
Hence, the paramagnetic contribution
$M_{p}$ to the total magnetisation originates from the zero-energy states
only, each of which contributes one Bohr magneton: $M_{p}=g\mu_{\rm B} $.
We note that this magnetisation can be written in terms of the Witten operator:
\begin{equation}
M_{p}=\mu_{\rm B} \Delta (\beta ).
\end{equation}
This relation, however, is no longer true if the spectrum of (4.2) contains
also a continuous part starting at $\varepsilon _{c}$, the so-called mobility
edge. That is, we assume a discrete spectrum with eigenvalues below
$\varepsilon _{c}$ and a continuous spectrum given by
$[\varepsilon _{c},\infty )$. Let us introduce another regularisation for
the trace of the Witten operator $P=\sigma _{3}$ leading to a modified Witten
index \cite{Comment3}:
\begin{equation}
\tilde\Delta (\varepsilon ):={\rm Tr\,}\Bigl(P\,\Theta (H-\varepsilon )\Bigr),
{}~~\varepsilon >0.
\end{equation}
Here $\Theta $ denotes the unit-step function
taking the value zero for negative arguments and the value one for positive
arguments. It is obvious that (4.6) agrees with the original Witten index if
$H$ has a pure point spectrum. However, in the case of a finite mobility
edge $\varepsilon _{c}<\infty $ we have:
\begin{eqnarray}
\tilde{\Delta}(\varepsilon )=g+\Theta (\varepsilon -\varepsilon _{c})
\int\limits_{\varepsilon_{c}}^{\varepsilon }\d E\,
\Bigl(\rho _{\uparrow}(E)-\rho _{\downarrow}(E)\Bigr),
\end{eqnarray}
where $\rho _{\uparrow}$ and $\rho _{\downarrow}$ denote the densities of the
spin-up and spin-down energy eigenstates, respectively.
The definition (4.6) allows to
express the magnetisation in terms of this modified Witten index:
\begin{equation}
M_{p}=\mu_{\rm B} \tilde{\Delta }(\varepsilon _{\rm F})
\end{equation}
where $\varepsilon _{\rm F}$ denotes the Fermi energy of the
two-dimensional non-interacting electron gas in the magnetic field $B$. Also
(4.8) is only exact for absolute zero temperature $T=0$, it will be good for
$k_{\rm B}T\ll \varepsilon _{\rm F}$ ($k_{\rm B}$: Boltzmann's constant).
Otherwise, the Fermi energy in (4.8) will have to be replaced by the
temperature-dependent chemical potential.

This result suggests that the existence and position $\varepsilon _{c}$
of a mobility edge for a certain magnetic field configuration
may be verified experimentally as
nowadays the Fermi energy in two-dimensional semiconductors can
be chosen rather freely in the preparation of the samples.
\subsection{The three-dimensional electron and the paramagnetic conjecture}
The above results on the two-dimensional electron can more or less be taken
over to the three-dimensional case. Here the supercharge may be defined as
follows \cite{CR83}:
\begin{equation}
Q':=\frac{1}{\sqrt{4m}}\sum_{i=1}^3\left[\left(p_{i}-\frac{e}{c}\,a_{i}(\vec{r})
\right)\otimes\sigma_{i}\right],
\end{equation}
where $\vec{r}:=(x,y,z)$ and $\vec{p}:=(p_{1},p_{2},p_{3})$ are the
position and momentum operators on ${\cal H}_{0}={\cal L}^2(\rz^3)$,
respectively. The corresponding Hamiltonian $H':=2Q'^2$
becomes the Pauli Hamiltonian of the three-dimensional electron. Let us note
that the Witten operator $P=\sigma _{3}$ does, in general, not commute with
$H'$. That is, the relations (2.2) are no longer true. Only in the special
case where the magnetic field
$\vec{B}(\vec{r}):=\vec{\nabla}\times\vec{a}(\vec{r})=B(x,y)\vec{e}_{z}$ is
translationary invariant along one axis, say the $z$-axis, all the results
found above remain valid. This is because of the relation $H'=H+p^2_{3}/2m$.
Hence, any subspace corresponding to a fixed eigenvalue of $p_{3}$ constitutes
a two-dimensional SUSY system as discussed in Section 4.1.

Nevertheless, the relation (4.8) between the
magnetisation and the modified Witten index is still
valid for arbitrary magnetic fields. However, the
Aharonov-Casher result cannot be taken over to three dimension. Indeed, here
it may be possible that SUSY is broken. For a pure point spectrum of $H'$ this
has the surprising consequence of a vanishing magnetisation as the Witten
index and its modified version is zero in this case.

Another consequence of the SUSY is that it provides a counterexample to the
paramagnetic conjecture due to Hogreve, Schrader and Seiler \cite{HSS78}. This
conjecture states that the groundstate energy of the general Pauli Hamiltonian
\begin{equation}
H_{\rm P}(\vec{a},V):=\frac{1}{2m}\left(\vec{p}-\frac{e}{c}\,\vec{a}\right)^2-
\mu_{\rm B}\left(\vec{\nabla}\times\vec{a}\right)\cdot\vec{\sigma }+V
\end{equation}
is always less or equal to that with zero magnetic field:
\begin{equation}
{\rm inf\,spec\,}\Bigl(H_{\rm P}(\vec{a},V)\Bigr)
\leq{\rm inf\,spec\,}\Bigl(H_{\rm P}(0,V)\Bigr).
\end{equation}
A proof of this inequality exist for arbitrary scalar potentials $V$ and
magnetic fields of the form $\vec{B}(\vec{r})=B(x,y)\vec{e}_{z}$ \cite{ASe79}.
However, in the general case Avron and Simon \cite{ASi79} found a
counter-example.
Nevertheless, it is believed, see ref.\ \cite{CFKS87} p.131,
that (4.11) ``$\ldots$ still holds for general $\vec{a}$ and selected sets
of $V \ldots$''. Here we note that for the particular set $\{V=0\}$ the
factorizability of $H_{\rm P}(\vec{a},0)\equiv H'=2Q'^2\geq 0$ implies
the inequality
\begin{equation}
{\rm inf\,spec\,}\Bigl(H_{\rm P}(\vec{a},0)\Bigr)
\geq{\rm inf\,spec\,}\Bigl(H_{\rm P}(0,0)\Bigr)=0.
\end{equation}
This inequality is in the opposite direction of the paramagnetic conjecture
(4.11). For arbitrary magnetic fields such that SUSY is a good symmetry the
inequality (4.12) can be replaced by an equality. However, for any magnetic
field which does break SUSY we have a strict inequality in (4.12) and thereby a
counterexample to the conjecture (4.11). Or vice versa, for any magnetic
field, for which the paramagnetic conjecture with $V=0 $ can be proven, SUSY
will be a good symmetry and hence equality holds in (4.12).
\ack
I would like to thank S.A.\ Baran and A.O.\ Barut for their kind invitiation
to participate in this conference. Part of the present work was done in
collaboration with A.\ Inomata. This collaboration has been supported by
the Deutsche Forschungsgemeinschaft which is greatfully acknowledged. Finally,
I thank A.\ Inomata and P.\ M\"uller for their remarks improving this
manuscript.
% References

%
 Figures and figure captions appended as postscript file.

\begin{thebibliography}{99}
\bibitem{SUSY1}H. Miyazawa, Phys.\ Rev.\ {\bf 170} (1968) 1586.\\
        Y.A.\ Gol'fand und E.P.\ Likhtman, JETP Lett.\ {\bf 13} (1971) 452.\\
        D.V.\ Volkov und V.P.\ Akulov, JETP Lett.\ {\bf 16} (1972) 621.\\
        J.\ Wess and B.\ Zumino, Nucl.\ Phys.\ {\bf B70} (1974) 39.
\bibitem{SUSY2}M.F.\ Sohnius, Phys.\ Rep.\ {\bf 128} (1985) 39.\\
P.G.O.\ Freund, {\it Introduction to Supersymmetry}, (Cambridge Univ.\ Press,
Cambridge, 1986).
\bibitem{KC85}V.A.\ Kosteleck\'{y} and D.K.\ Campbell eds.,
{\it Supersymmetry in Physics}, (North-Holland, Amsterdam, 1985).
\bibitem{N76}H.\ Nicolai, J.\ Phys.\ {\bf A 9} (1976) 1497.
\bibitem{W81}E.\ Witten, Nucl.\ Phys.\ {\bf B188} (1981) 513.
\bibitem{W82}E.\ Witten, Nucl.\ Phys.\ {\bf B202} (1982) 253.
\bibitem{CFKS87} H.L. Cycon, R.G.\ Froese, W.\ Kirsch and B.\ Simon, {\em
Schr\"odinger Operators with Application to Quantum Mechanics and Global
Geometry}, (Springer, Berlin, 1987).
\bibitem{T92}B.\ Thaller, {\it The Dirac Equation}, (Springer, Berlin, 1992).
\bibitem{Comment1}This regularization has been suggested by Witten \cite{W82}.
However, it should be noted that the Witten index (2.9) is only well defined
if the operator $P\e^{-\beta H}$ is trace class.
For a continuous spectrum of $H$ even some more requirements have to be
imposed on this operator. In the latter case one may start with a
finite configuration space such as a $d$-dimensional box of volume $L^{d}$,
$L>0$, and, after the evaluation of the trace (2.9), take (if possible) the
limit $L\to\infty $ in an appropriate way.
\bibitem{FC85}B.\ Freedman and F.\ Cooper, Physica {\bf 15D} (1985) 138.
\bibitem{IJ92}A.\ Inomata and G.\ Junker, in H.A.\ Cerdeira, S.\ Lundqvist, D.\
Mugnai, A.\ Ranfagni, V.\ Sa-yakanit and L.S.\ Schulman eds., {\it Lectures on
Path Integration: Trieste 1991}, (World Scientific, Singapore, 1993) p.460.
\bibitem{S85}C.V.\ Sukumar, J.\ Phys.\ {\bf A 18} (1985) 2917.
\bibitem{IJ93}A.\ Inomata and G.\ Junker, in L.Q.\ Liang, M.L.\ Wang, S.N.\
Qiao and D.C.\ Su eds., {\it Proceedings of International Symposium on
Advanced Topics in Quantum Physics}, (Science Press, Beijing, 1993) p.61.
\bibitem{IJ94}A.\ Inomata and G.\ Junker, {\it Quasi-classical path integral
approach to supersymmetric quantum mechanics}, preprint (1994).
\bibitem{E86}B.\ Eckhardt, Phys.\ Lett.\ {\bf 168B} (1986) 245.
\bibitem{IJS93}A.\ Inomata, G.\ Junker and A.\ Suparmi, J.\ Phys.\ {\bf A 26}
(1993) 2261.
\bibitem{CBC85}A.\ Comtet, A.D.\ Bandrauk and D.K.\ Campbell, Phys.\ Lett.\
{\bf 150B} (1985) 159.
\bibitem{G83}L.\'{E}.\ Gendenshte\^{i}n, JETP Lett.\ {\bf 38} (1983) 356.
\bibitem{IH51}L.\ Infeld and T.E.\ Hull, Rev.\ Mod.\ Phys.\ {\bf 23} (1951) 21.
\bibitem{S92}A.\ Suparmi, {\it Doctoral thesis}, SUNY-Albany (1992).
\bibitem{DGKPS93}R.\ Dutt, A.\ Gangopadyaya, A.\ Khare, A.\ Pagnamenty and
U.\ Sukhatme, Phys.\ Lett.\ {\bf A 174} (1993) 363.
\bibitem{BR92}A.O.\ Barut and P.\ Roy, in A.\ Frank, T.H.\ Seligman and K.B.\
Wolf eds., {\it Group theory in physics}, AIP Conference Proceedings 266,
(AIP, New York, 1992), p.248.
\bibitem{UBK94}C.A.\ Uzes, A.O.\ Barut and E.\ Kapuscik, {\it From an Infinite
to a Finite
Number of Degrees of Freedom}, Paper presented in this conference.\\
E.\ Kapuscik, C.A.\ Uzes and A.O.\ Barut, {\it Quantization of
constraint solutions}, Phys.\ Rev.\ {\bf A 49} (1994) in print.\\
I would like to thank C.A.\ Uzes for providing me copies of these articles
prior to publication.
\bibitem{N91}H.\ Nicolai, Phys.\ Bl.\ {\bf 47} (1991) 387 (in German).
\bibitem{CR83} M.\ deCrombrugge and V.\ Rittenberg, Ann.\ Phys.\ (NY) {\bf
151} (1983) 99.
\bibitem{K38}H.A.\ Kramers, {\it Die Grundlagen der Quantentheorie:
Quantentheorie des Elektrons und der Strahlung}, (Akademische
Verlagsgesellschaft, Leipzig, 1938).\\
I am thankful to A.O.\ Barut for drawing my attention to this reference.
\bibitem{AC78} Y.\ Aharonov and A.\ Casher, Phys.\ Rev.\ {\bf A 19} (1978)
1461.
\bibitem{Comment2} Such a case is an example where the integral in (4.3)
diverges and the degeneracy of the groundstate is infinite. In all these cases
one should consider specific quantities. In other words, one uses a
regularisation as suggested in \cite{Comment1} and considers the degeneracy
per unit area $\lim_{L\to\infty }(g/L^2)$. A similar procedure is also
mandatory for the evaluation of the magnetisation (4.5) and (4.8),
respectively.
\bibitem{Comment3}The comment \cite{Comment1} made for the Witten index (2.9)
also applies here.
\bibitem{HSS78} H.\ Hogreve, R.\ Schrader and R.\ Seiler, Nucl.\ Phys.\ {\bf
B142} (1978) 525.
\bibitem{ASe79} J.E.\  Avron and R.\ Seiler, Phys.\ Rev.\ Lett.\ {\bf 42}
(1979) 931.
\bibitem{ASi79} J.\  Avron and B.\ Simon, Phys.\ Lett.\ {\bf 75A} (1979) 41.
\end{thebibliography}
\end{document}